\documentclass[reprint, pre, aps]{revtex4-2}
\usepackage{soul}
\usepackage{amsmath}
\usepackage{amssymb}
\usepackage{mathtools}
\usepackage{graphicx}
\usepackage{booktabs}
\usepackage{longtable}
\usepackage{bm}
\usepackage[version=4]{mhchem}
\usepackage[colorlinks=true,citecolor=blue,linkcolor=blue]{hyperref}
\begin{document}

\title{Circular polarization of fast radio bursts by asymmetric erosion in longitudinally magnetized plasma}
\author{Da-Chao Deng}
\author{Hui-Chun Wu}
\email{huichunwu@zju.edu.cn}
\affiliation{Institute for Fusion Theory and Simulation, School of Physics, Zhejiang University, Hangzhou 310058, China}%
\date{\today}
\begin{abstract}
  Magnetars are likely to be the origin of all fast radio bursts. The recent detection of circularly polarized bursts suggests that they might be generated deep inside magnetar magnetospheres. However, the mechanism behind the circular polarization remains uncertain. Here, we study the propagation of an intense radio pulse in a longitudinally magnetized electron-dominant plasma by particle-in-cell simulations. When the field strength of the radio pulse exceeds the background magnetic field, it can excite a nonlinear plasma wakefield and continually erodes due to energy transfer to the wake. Along the magnetic field, the plasma wakefield launched by the right-circularly polarized pulse is much stronger and more nonlinear than that by the left-circularly polarized pulse. Hence, the erosion rates of the two circularly polarized modes are significantly different. We discover that this asymmetric erosion can generate circularly polarized modes from a linearly polarized pulse at relativistic intensities, even when the cyclotron frequency is much higher than the radio frequency. Finally, we present a proof-of-principle simulation to demonstrate the generation of circularly polarization by this magneto-induced asymmetric erosion in the nonuniform environment of magnetar magnetospheres.
\end{abstract}

\maketitle

\section{Introduction}
Fast radio bursts (FRBs), highly intense radio transients lasting only milliseconds, remain one of the universe's great mysteries \cite{zhangPhysicsFastRadio2023,xiaoPhysicsFastRadio2021}. Since the detection of FRB 20200428D from a Galactic magnetar SGR J1935+2154 \cite{andersenBrightMilliseconddurationRadio2020}, magnetars have been widely accepted to be one source of FRBs. However, there is no consensus on the radiation mechanism of FRBs. The close-in models suggest that FRBs originate within or nearby magnetar magnetospheres \cite{kumarFastRadioBurst2017,quPolarizationFastRadio2023}, while the faraway models argue that they are generated far outside through relativistic shocks driven by magnetar winds \cite{lyubarskyModelFastExtragalactic2014,beloborodovBlastWavesMagnetar2020,iwamotoLinearlyPolarizedCoherent2024}.

The observed polarization properties disfavor the faraway model. Most FRBs are highly linearly polarized (LP) \cite{michilliExtremeMagnetoionicEnvironment2018,choSpectropolarimetricAnalysisFRB2020,dayHighTimeResolution2020}, and both types of models are capable of accounting for this fact \cite{quPolarizationFastRadio2023,iwamotoLinearlyPolarizedCoherent2024}. Notably, several LP bursts from FRB 20180301 are observed to possess various polarization angle swings. These swing patterns can be produced through the sweeping of the line of sight or the change of the magnetospheric configurations, but are difficult to generate in the faraway models \cite{luoDiversePolarizationAngle2020}. In addition, some bursts display circularly polarized (CP) degrees ranging from 10\% to 75\% \cite{choSpectropolarimetricAnalysisFRB2020,dayHighTimeResolution2020,hilmarssonPolarizationPropertiesFRB2021,kumarCircularlyPolarizedRadio2022,fengCircularPolarizationTwo2022}. Recently, CP bursts were detected with a 90\% CP degree from the repeating source FRB 20201124A \cite{jiangNinetyPercentCircular2025}. For the faraway models, the only known way to generate CP modes is through burst propagation in ambient plasmas, which generally requires relatively strong magnetic fields. However, a nonmagneto-ionic environment has been suggested for FRB 20220912A, from which some bursts show 70\% CP degrees \cite{fengExtremelyActiveRepeating2024}.

It is highly probable that some propagation effects inside magnetar magnetospheres might produce the CP pulses. For the close-in models, FRBs are likely generated on the magnetic polar cap and travel out from the open-field-line region \cite{luUnifiedPictureGalactic2020,quTransparencyFastRadio2022}. Thus, throughout this paper, we focus on the scenario that the pulses propagate along the background magnetic field $\bm{B}_\mathrm{bg}$.

In this context, an asymmetry between electrons and positrons, such as differing spatial distributions or streaming momenta, is generally necessary for a pair plasma to respond differently to the left- and right-circularly polarized (LCP and RCP) components. Notably, a few particle-in-cell (PIC) simulations of pulsar or magnetar magnetospheres have demonstrated high asymmetries of densities or momenta of electrons and positrons, particularly relevant to discharge activities \cite{timokhinPOLARCAPCASCADE2015,philippovINITIOPULSARMAGNETOSPHERE2015}. More recently, several pairs of LCP and RCP nanoshots with a uniform interval $21\;\mu \mathrm{s}$ were identified from a giant pulse emitted along the magnetic axis of Crab pulsar \cite{wuMechanismCircularPolarization2024}. The extreme Faraday effect in a highly asymmetric plasma has been adopted to explain these CP nanoshot pairs. Therefore, highly asymmetric pair plasma are likely to exist in magnetospheres.

One of the widely discussed CP mechanisms is cyclotron absorption \cite{luoCyclotronAbsorptionRadio2001,wangPolarizationChangesPulsars2010,daiCircularPolarizationRepeating2021}, which can damp the LCP (or RCP) component when the cyclotron frequency $\Omega$ of the electrons (or positrons) equals the Doppler-shifted frequency $\omega'$ of the electromagnetic wave in the rest frame of the electrons (or positrons). When the damping of one CP component dominates, the residual part forms the observed CP bursts. However, the resonant condition $\Omega=\omega'$ applies only to linear plasma theory \cite{goldstonIntroductionPlasmaPhysics1995}, and is also hardly fulfilled due to the strong magnetic fields in magnetospheres. On the other hand, since the field strength of these radio waves and the magnetic field of neutron stars attenuate inversely with the first and third powers of distance, respectively, the radio-wave field $E_0$ will exceed the neutron-star magnetic field $B_\mathrm{bg}$ at a certain distance, causing relativistic dynamics induced by radio waves to dominate. In this nonlinear regime, strong coupling between radio waves and plasmas makes it unclear whether and how two CP components independently respond to or act on plasmas. This demands a self-consistent investigation on the collective plasma behavior under strong LP waves.

In this paper, we use one-dimensional particle-in-cell simulations to study the nonlinear propagation of intense radio waves along the magnetic field in an electron-dominated plasma. In the unmagnetized case, a plasma density spike, pushed up by the large ponderomotive force, continually erodes the pulse front \cite{bulanovNonlinearDepletionUltrashort1992, deckerEvolutionUltraintenseShortpulse1996}. The LCP and RCP pulses are shown to have the same erosion rates due to parity symmetry. However, a longitudinal magnetic field $B_\mathrm{bg}$ can enhance the wake field launched by the RCP pulse and suppress that by the LCP pulse, hence the symmetry is broken. We show that the significantly different erosion rates of both CP modes enable the generation of one CP mode from an intense LP pulse. The application of this magneto-induced asymmetric erosion (MIAE) to the CP modes of FRBs is verified by a proof-of-principle simulation in magnetospheres.

\section{Simulation setups}
Most FRBs in GHz exhibit narrow bandwidths $\Delta \nu \sim 100\;\mathrm{MHz}$, implying a short coherent time $1/\Delta \nu \sim 10\;\mathrm{ns}$ \cite{zhangUpperFieldstrengthLimit2022,wuMechanismCircularPolarization2024}. Observationally, some millisecond FRBs have been resolved to contain subpulses lasting from $60\;\mathrm{ns}$ to $50\;\mathrm{\mu s}$ \cite{choSpectropolarimetricAnalysisFRB2020,dayHighTimeResolution2020,nimmoHighlyPolarizedMicrostructure2021,nimmoBurstTimescalesLuminosities2022}. Hence, we focus on propagation of these nanosecond bursts in plasma.

The PIC code used here is EPOCH \cite{arberContemporaryParticleincellApproach2015a}. The simulation window spans $12\;\mathrm{m}$ in length with 4000 grids. A moving window at the speed of light $c$ is used to simulate long-distance propagation. The electric field of the radio waves is given by $\bm{E}=E_0\exp[ -(t-t_0-x/c)^2/T^2][ \sin[\omega (t-x/c)]\bm{e}_y-\delta\cos[\omega (t-x/c)]\bm{e}_z ]$, where $E_0$ is the wave amplitude. We take the frequency $\omega/2\pi = 1\;\mathrm{GHz}$, the pulse duration $T=6\tau_0$, and the time offset $t_0=18\tau_0$, where $\tau_0=1\;\mathrm{ns}$ is the wave cycle. The factor $\delta$ denotes the polarization:  $-1$ for LCP, $1$ for RCP, and $0$ for LP. We normalize the pulse amplitude and background magnetic field as $a_0=eE_0/m_\mathrm{e}\omega c\approx \frac{E_0}{357.24\;\mathrm{statV/cm}}$ and $b=eB_\mathrm{bg}/m_\mathrm{e}\omega c\approx \frac{B_\mathrm{bg}}{357.24\;\mathrm{Gs}}$, where $m_\mathrm{e}$ is the electron mass and $e$ is the elementary charge. The background magnetic field is along the $x$-axis.

To simulate highly asymmetric pair plasmas, we simply set up a pure electron-ion plasma. In the magnetosphere, an unscreened parallel electric field accelerates electrons and positrons in opposite directions, creating a streaming plasma. Physically, the extent of any propagation effect is determined mainly by the relative cross distance between the radio waves and the plasma. Compared to wave propagation in stationary or counterstreaming species, the net crossing between co-moving waves and plasma/particles is significantly slower \cite{wuMechanismCircularPolarization2024}. Thus, the response of electrons could dominate as they collide head-to-head with the radio bursts. The normalized pulse amplitude $a_0$ is Lorentz invariant. In the electron rest frame, the relativistic increase in positron mass is significant, hence we roughly use stationary ions to replace the streaming positrons. The initial plasma has a uniform density with the left boundary at $x=30\lambda$, where $\lambda \approx 30\;\mathrm{cm}$ is the wavelength. There are 25 macroparticles per cell. Plasma density is given in units of the critical density $n_\mathrm{c}=m\omega^2/4\pi e^2\approx 1.24 \times 10^{10}\;\mathrm{cm}^{-3}$.

\begin{figure}[tp]
  \centering
  \includegraphics[scale=0.487]{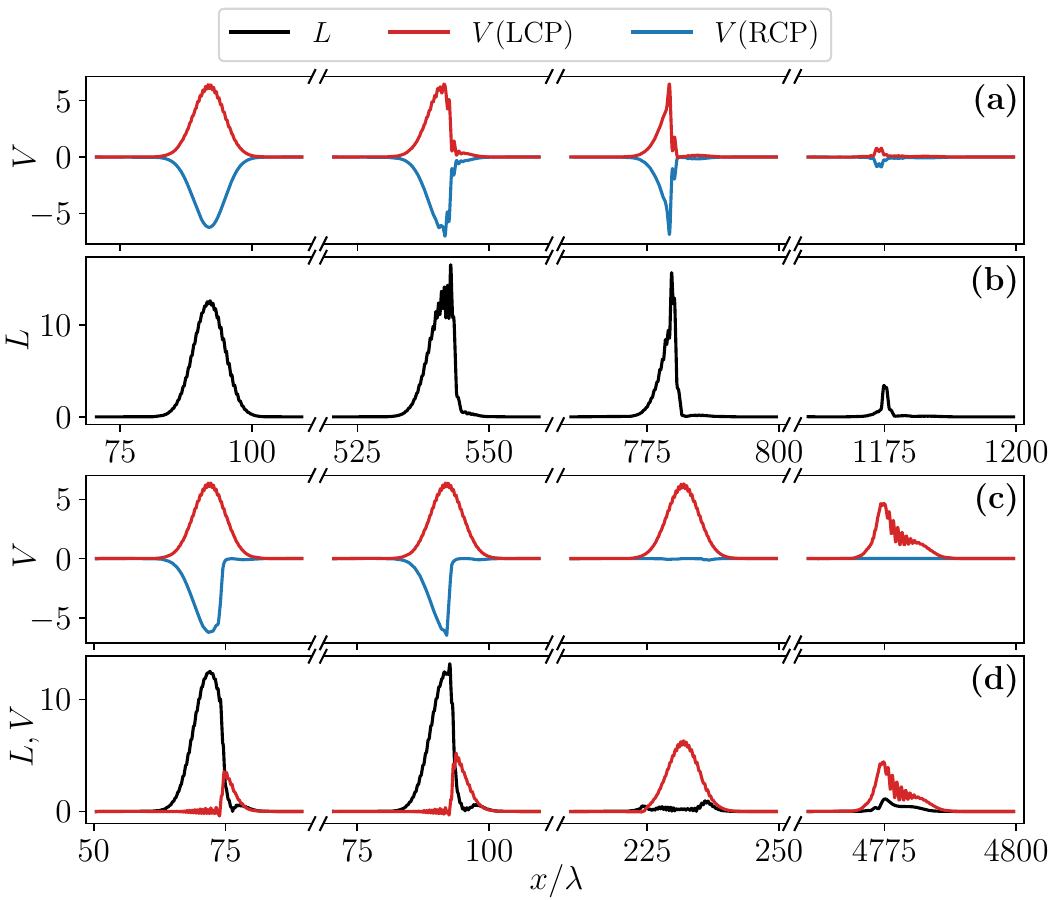}
  \caption{Erosion of intense radio pulses in plasmas. (a) LCP, RCP and (b) LP pulses in the unmagnetized plasmas at $t/\tau_0=110$, $560$, $800$, and $1200$. (c) LCP, RCP and (d) LP pulses in the magnetized plasmas with $b=1.5$ at $t/\tau_0=90$, $110$, $800$, and $4800$. The intensity profiles of CP and LP pulses are given by the Strokes parameters $V$ and $L$ ($V>0$ for LCP, $V<0$ for RCP, and $L$ for LP). We take $a_0=2.5$ for CP, $a_0=5$ for LP, and $n_0=0.01n_\mathrm{c}$.}
  \label{fig1}
\end{figure}

\section{Magneto-induced asymmetric erosion}
Firstly, we discuss the propagation of intense radio waves in an unmagnetized plasma with initial density $n_0=0.01n_\mathrm{c}$. Figure~\ref{fig1}(a) presents the intensity profiles of the LCP and RCP pulses with $a_0=2.5$. The LP case at $a_0=5$ is shown in Fig.~\ref{fig1}(b). The energy of the LP pulse is equal to the sum of the two CP pulses. The intensity envelopes with different polarizations are calculated by the Stokes parameters \cite{goldsteinPolarizedLight2011} using the normalized pulse amplitude $a_0$, where $V>0$ for LCP, $V<0$ for RCP, and $L$ for LP. It is found that all of these pulses undergo continuous erosion of their fronts. Due to parity symmetry \cite{jacksonClassicalElectrodynamics1999}, both CP pulses exhibit almost the same results.

\begin{figure}[t]
  \centering
  \includegraphics[scale=0.487]{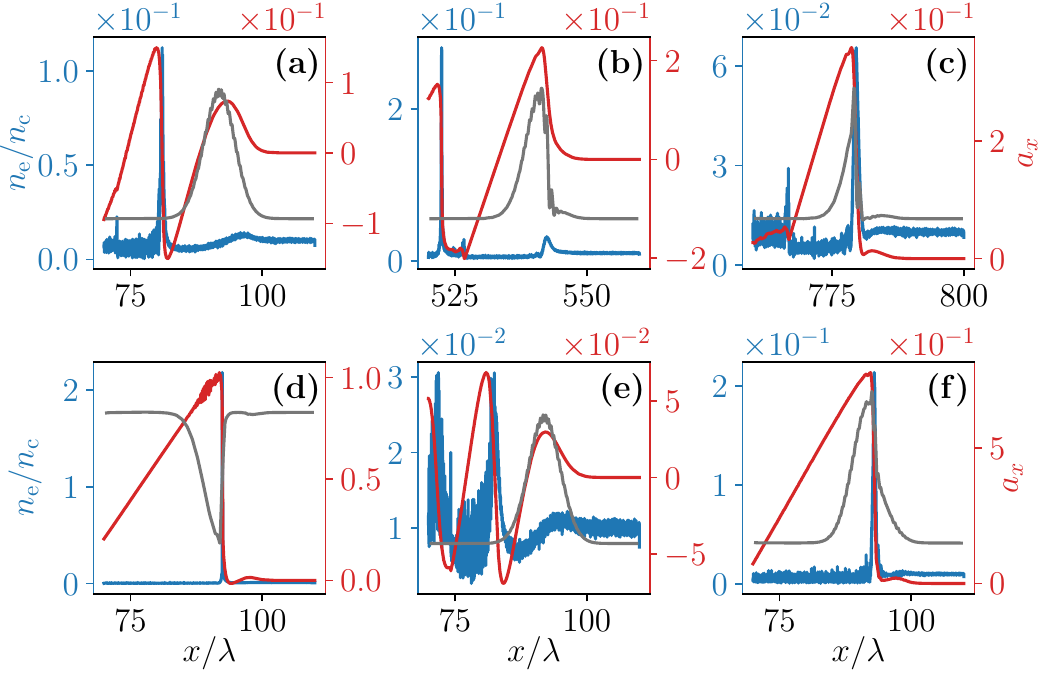}
  \caption{Plasma wakefields excited by radio pulses. (a-c) Snapshots of the electron density and longitudinal electric field excited by the LCP pulse in the unmagnetized plasma [Fig.~\ref{fig1}(a)] at $t/\tau_0=110$, $560$, and $800$. Snapshots of the wakefields by (d) RCP, (e) LCP, and (f) LP pulses at $t/\tau_0=110$ in the magnetized plasmas with $b=1.5$ [Fig.~\ref{fig1}(c,d)]. The intensity profiles (gray) are plotted in arbitrary units.}
  \label{fig2}
\end{figure}

Plasma dynamics induced by the LCP pulse are presented in Figs.~\ref{fig2}(a-c), which show the snapshots of the electron density  $n_\mathrm{e}$ and the longitudinal electric field $a_x = eE_x/mc\omega$. The intense pulse first piles up the electrons at its front and then accelerates them forward. When the space-charge field becomes stronger than the ponderomotive force, the electrons are pulled back, forming a wakefield. Thus, the pulse is depleted due to the energy transfer to the wake \cite{deckerEvolutionUltraintenseShortpulse1996}.

For $a_0>1$, the plasma wakefield is nonlinear, manifested by a spike-shaped density distribution and an elongated plasma wavelength \cite{esareyPhysicsLaserdrivenPlasmabased2009}. The nonlinearity of the wake results in the depletion being localized at the pulse front, steepening its leading edge \cite{deckerEvolutionUltraintenseShortpulse1996}. As the pulse erodes backward toward its central peak at $t=800\tau_0$ [Fig.~\ref{fig2}(c)], the excited wakefield becomes more nonlinear due to the maximized ponderomotive force, which is proportional to the gradient of the pulse intensity. Meanwhile, the density peaks rise, the intervals between them widen, and the space-charge field strengthens. The maxima of the density peak and the wakefield can reach $n_\mathrm{e}\approx 0.065n_\mathrm{c}$ and $a_x\approx 0.358$, respectively.

\begin{figure}[tb]
  \centering
  \includegraphics[scale=0.487]{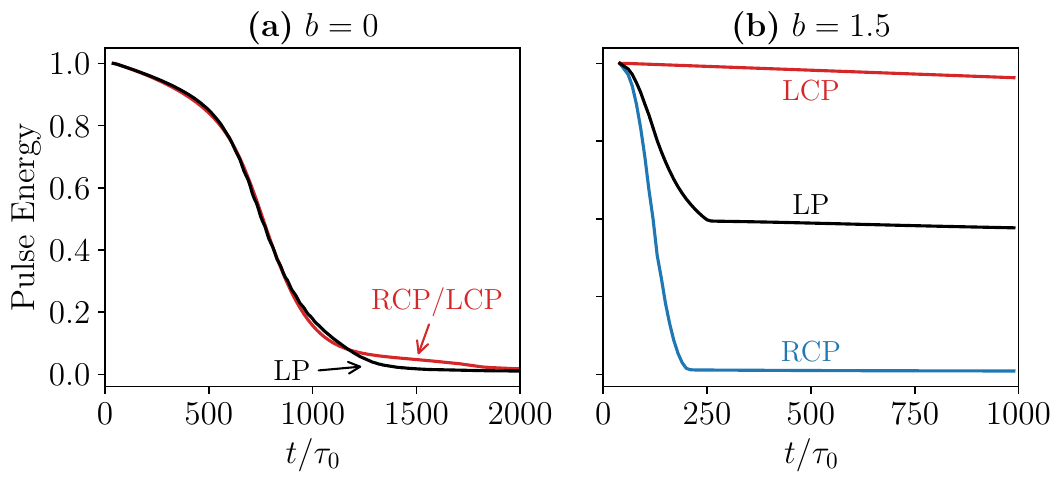}
  \caption{Energy evolution with time for the (a) unmagnetized [Fig.~\ref{fig1}(a-b)] and (b) magnetized [Fig.~\ref{fig1}(c-d)] cases. The energies are all normalized to the initial pulse energies. }
  \label{fig3}
\end{figure}

Figure~\ref{fig3}(a) shows the pulse energies as a function of time. All CP and LP waves undergo almost the same erosion processes and are exhausted at $t\approx 2000\tau_0$. Although the LP pulse contains twice the energy of one CP pulse, the simulation shows the wakefield intensity $E_x^2$ of the LP pulse is also twice of the CP case. Thus, they have the same erosion rates.

A longitudinal background magnetic field $\bm{B}_\mathrm{bg}$ can break the parity symmetry of the two CP modes. Figure~\ref{fig1}(c) shows the evolutions of the LCP and RCP pulses for $b=1.5$. Compared to the unmagnetized case [Fig.~\ref{fig1}(a)], the RCP pulse in the magnetized plasma forms a sharper front and erodes more quickly. In contrast, the LCP pulse exhibits the opposite trend, forming a gentle slope at its front and eroding more slowly. This discrepancy occurs because the longitudinal magnetic field significantly enhances (or suppresses) the wakefield excited by the RCP (or LCP) pulse. Figures~\ref{fig2}(d,e) display the wakefields at the same moment $t=110\tau_0$ for the RCP and LCP pulses, respectively. Due to the faster depletion, the RCP pulse has eroded to its center earlier, pushing up a high-density spike of $n_\mathrm{e}\approx 2n_\mathrm{c}$. The corresponding wakefield peaks at $a_x\approx 1$, which is 2.8 times that of the unmagnetized case [Figs.~\ref{fig2}(c)]. However, for the LCP pulse in Fig.~\ref{fig2}(e), the wakefield is significantly suppressed by $B_\mathrm{bg}$, resulting in a much slower erosion.

The effect of the longitudinal magnetic field can be understood using a single-electron picture. As an RCP pulse propagates along a static magnetic field, an electron initially at rest will rotate in the same direction as the rotating electric field of the pulse and gain energy from the field. Its kinetic energy varies with $E_\mathrm{k}/m_\mathrm{e}c^2=a_0^2/(1-b)^2$ \cite{salaminExactAnalysisUltrahigh2000}, which indicates the resonant condition $b=\omega/\Omega=1$, the same as that in linear plasma theory. For a fixed $a_0$, when $b \rightarrow 1$, the electron energy increases, manifesting more nonlinear motion. However, for a group of electrons in plasma, the resonant condition is modified to $b/a_0\approx 1$ due to relativistic effects. Therefore, when $b\gg 1$, the intense field will bring the wave-plasma interaction close to resonance, leading to more nonlinear plasma dynamics. For an LCP pulse, the electron kinetic energy has $E_\mathrm{k}/m_\mathrm{e}c^2=a_0^2/(1+b)^2$, and hence the electron motion is always constrained by the magnetic field $b$.

Now, we examine the important LP case with $a_0=5$ and $b=1.5$ in Fig.~\ref{fig1}(d). Despite the strongly coupled nonlinearity of the wakefield, the erosion process of this intense LP driver can still be understood as the independent erosions of two CP modes, as clearly identified by comparing Figs.~\ref{fig1}(c) and \ref{fig1}(d). Because of the asymmetric erosion caused by $B_\mathrm{bg}$, the RCP component in the LP pulse erodes faster, leaving an almost undeformed LCP pulse. The RCP component is competely exhausted at $t\approx 250\tau_0$, and then the erosion process becomes nearly identical to a single LCP pulse. Before the complete erosion of the RCP component [Fig.~\ref{fig2}(f)], the density spike always resides at the site of its erosion, where the energy transfer dominates.

As shown in Fig.~\ref{fig3}(b), the steep (or gentle) slope in the LP erosion curve corresponds to faster (or slower) erosion of the RCP (or LCP) component. The full depletion of the RCP component in the LP pulse occurs slightly later than in the single RCP pulse, because the wakefield amplitude in Fig.~\ref{fig2}(f) excited by the LP pulse falls intermediate between the RCP and LCP pulses.

\section{Impact of parameters}
In the following, we will discuss the influence of changing magnetic field and plasma density on the above results. For $b<a_0$, we find that a stronger magnetic field can further increase the erosion rate of the RCP pulse, and meanwhile decrease that of the LCP pulse, thereby intensifying the asymmetric erosion between two CP modes. This trend aligns with the understanding of the nonlinear relativistic-induced resonance. For $b\gg a_0$, the plasma response is essentially linear due to strong magnetic confinement, and the wakefields of both CP modes are so weak that the pulse erosion almost ceases.

\begin{figure}[t]
  \centering
  \includegraphics[scale=0.487]{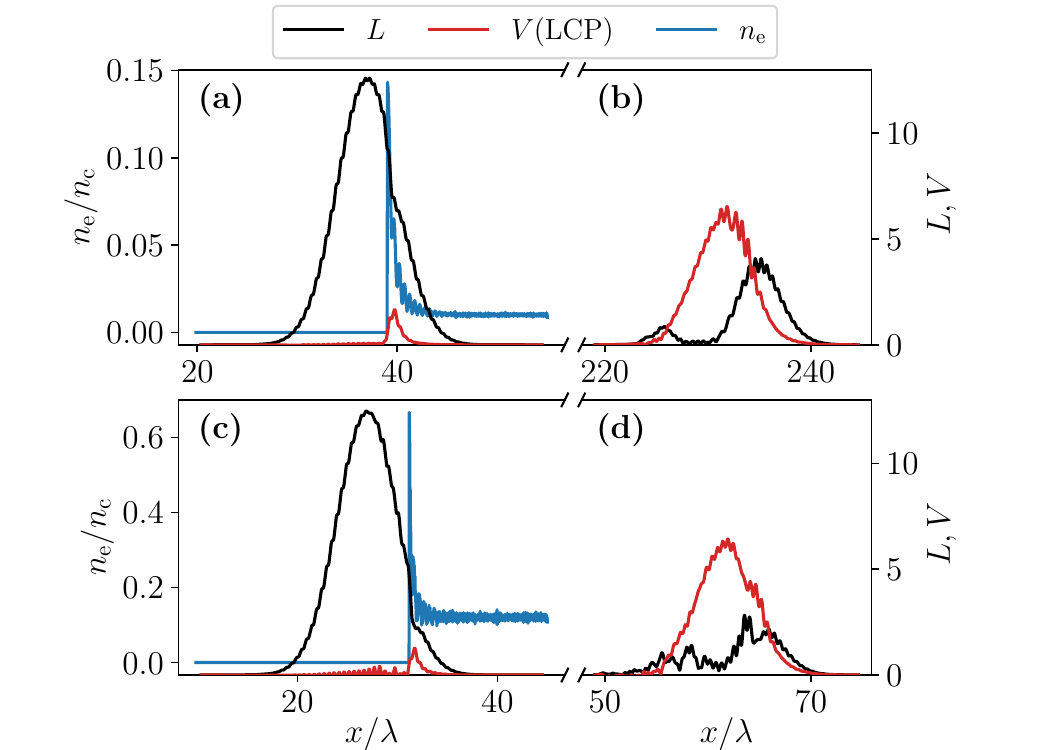}
  \caption{Erosion of the LP pulse at a higher magnetic field and density. (a,b) Snapshots at $t/\tau_0=55$ and $250$ for $a_0=b=2.5$ and $n_0=0.01n_\mathrm{c}$. (c,d) Snapshots at $t/\tau_0=45$ and $80$ for $a_0=b=2.5$ and $n_0=0.12n_\mathrm{c}$. }
  \label{fig4}
\end{figure}

Nearby the resonance at $b \approx a_0$, the key finding is that the erosion of the RCP component starts in the middle of its profile. Figures~\ref{fig4}(a,b) show the asymmetric erosion of an LP pulse with $a_0=b=2.5$. The erosion of the RCP component occurs around $2.31\lambda$ ahead of the pulse center at $t\approx 55\tau_0$, and then proceeds backward, damping almost all of the RCP energy behind the initial erosion point at $t\approx 250\tau_0$. Ultimately, the pulse’s front part before the initial erosion point remains LP while the tail portion exhibits LCP due to the absence of the RCP component. If we apply a higher $b$, the initial erosion point will shift further toward the pulse tail, until no erosion occurs in the entire pulse, i.e., the process enters the linear regime of $b\gg a_0$.

The initial plasma density $n_0$ can also affect the erosion rates of both CP components and the initial erosion point of the RCP component. Figures~\ref{fig4}(c,d) present the LP case with $b=2.5$ at a higher density of $n_0=0.12n_\mathrm{c}$. It can be observed that the depletion of the RCP component is complete at $t=80\tau_0$, much earlier than at $250\tau_0$ in the low-density case of $0.01n_\mathrm{c}$ [Fig.~\ref{fig4}(b)]. With the same travel distance, a higher density will involve more particles in the energy transfer, which increases the erosion rates of both CP modes and results in a higher density spike at the erosion point. The electron spikes in Figs.~\ref{fig4}(a) and \ref{fig4}(c) have densities of $0.14n_\mathrm{c}$ and $0.7n_\mathrm{c}$, respectively. In addition, it is notable that the initial erosion point shifts a bit forward for the high-density case. As a result, the CP degree in Fig.~\ref{fig4}(d) is greater than that in Fig.~\ref{fig4}(b), i.e., 88.6\% versus 77.1\%.

In short, a stronger magnetic field will shift the initial erosion point of the RCP component from the head to the tail in the LP pulse envelope, while a higher plasma density will make it shift in the opposite direction. The interplay of these two parameters influences the final CP degrees, and meanwhile leads to an interesting phenomenon as follows.

\section{For magnetar magneotospheres}
The depicted MIAE effect for generating CP modes should apply to magnetar magnetospheres. The background magnetic field in the magnetosphere is described as $B_\mathrm{bg}=B_\mathrm{s}(r/r_0)^{-3}$ \cite{lorimerHandbookPulsarAstronomy2005}, where $B_\mathrm{s}$ is the surface magnetic field, $r_0$ is the magnetar's radius, and $r$ is the distance from the magnetar. By $E_0\approx(L/r^2c)^{1/2}$, where $L$ is the isotropic luminosity, we obtain the ratio
\begin{equation}
  \begin{split}
    \frac{a_0}{b}
    & =\frac{E_0}{B_\mathrm{bg}}                                                                                                                                                                     \\
    & \approx5.78 \times 10^7 \bigg( \frac{r}{1\;\mathrm{km}} \bigg)^2\bigg( \frac{L}{10^{42}\;\mathrm{erg/s}} \bigg)^\frac{1}{2}\bigg( \frac{B_\mathrm{s}}{1\;\mathrm{Gs}} \bigg)^{-1},
  \end{split} \label{eq:L}
\end{equation}
which is proportional to $r^2$ and reaches its maximum at the magnetosphere boundary defined by the light cylinder radius $r_\mathrm{LC}=cP/2\pi$. Here, $P$ is the magnetar's spin period. Based on the above simulation results, we anticipate that the MIAE effect can occur inside the magnetosphere if $a_0/b>1$ holds at $r=r_\mathrm{LC}$. We take $r_0=10\;\mathrm{km}$, $P=1\;\mathrm{s}$, and $B_\mathrm{s}=10^{14}\;\mathrm{Gs}$ for a typical magnetar. For the weakest FRB with $L\approx10^{38}\;\mathrm{erg/s}$ \cite{zhangPhysicsFastRadio2023}, it follows that $a_0/b\approx 13$ at $r=r_\mathrm{LC}\approx 48000\;\mathrm{km}$, and $a_0$ is greater than $b$ over a distance of up to $34000\;\mathrm{km}$. Thus, the nonlinear propagation of FRBs should be taken into account in a significant portion of the magnetosphere.

The magnetosphere parameters are too extreme to simulate directly. The rotating magnetar can induce an electric field that separates charged particles, forming a net Goldreich-Julian charge density $\rho_\mathrm{GJ}\approx -\bm{\Omega}_\mathrm{m}\cdot \bm{B}_\mathrm{bg}/2\pi c$, where $\bm{\Omega}_\mathrm{m}$ is the angular velocity of the magnetar \cite{lorimerHandbookPulsarAstronomy2005}. The plasma density in the magnetar magnetosphere is thus $n_\mathrm{GJ}=\rho_\mathrm{GJ}/e\approx 6.9 \times 10^{10}\;\mathrm{cm}^{-3}\big( \frac{B_\mathrm{bg}}{10^{12}\;\mathrm{Gs}} \big)\big( \frac{P}{1\;\mathrm{s}} \big)^{-1}$. The actual electron and positron densities can be larger, with a multiplication factor $M$ ranging from $10^3$ to $10^5$ \cite{timokhinPOLARCAPCASCADE2015}. We then obtain $b=B_\mathrm{bg}e/m\omega c \approx 1.04 \times 10^{25}(r/\lambda)^{-3}$, and $n/n_\mathrm{c}\approx 2.06 \times 10^{19}(r/\lambda)^{-3}$ for $M=10^3$, both of which, when deep inside the magnetospheres, are too large for PIC simulations.

Nevertheless, by noticing that $n/n_\mathrm{c}\propto (r/\lambda)^{-3}$ and $h \equiv bn_\mathrm{c}/n$ is a constant, we perform a proof-of-principle simulation by modeling $n_0/n_\mathrm{c}=\bar{n}\lambda^3/(x+\Delta x)^{3}$ and $b=hn_0/n_\mathrm{c}$, where $\bar{n}$ and $\Delta x$ adjust the plasma density distribution. Here, we inject an LP pulse with $a_0=100$ from the left boundary, and choose $b=80$ and $n_0=0.02n_\mathrm{c}$ at the plasma surface. To simulate the nonuniform magnetospheric environment, we adopt $h=4000$, $\bar{n}\approx7.9 \times 10^9$, and $\Delta x=6500\lambda$. Moreover, we double the resolution of the simulation to resolve the cyclotron motion in such a strong magnetic field.

\begin{figure}[t]
  \centering
  \includegraphics[scale=0.487]{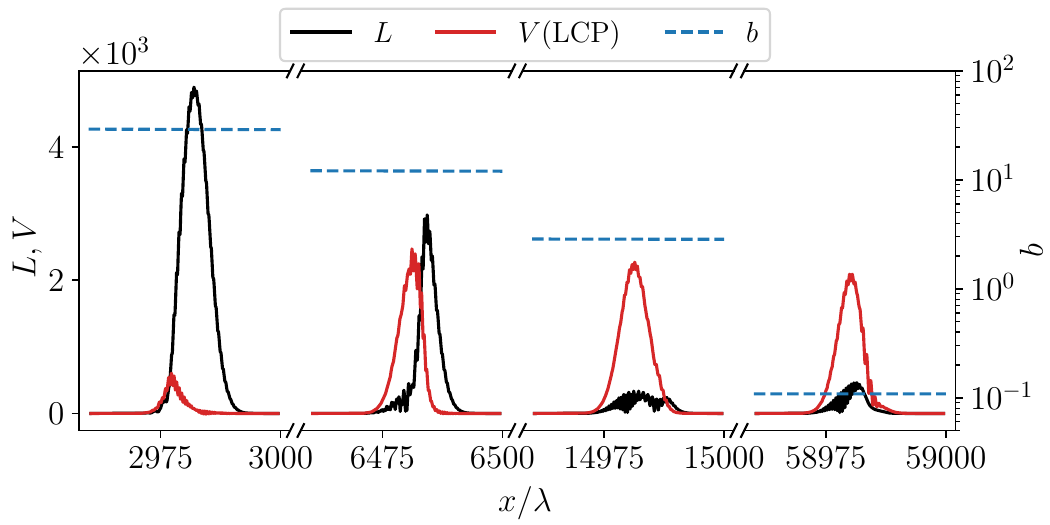}
  \caption{Evolution of ultra-intense LP pulse in the magnetosphere at $t/\tau_0=3000$, $6500$, $15000$, and $59000$. The blue dashed lines display the variation in the local background magnetic field. We take $a_0=100$, $n_0/n_\mathrm{c}\approx \bar{n}\lambda^3/(x+\Delta x)^{3}$, and $b=hn_0/n_\mathrm{c}$, where $\bar{n}\approx7.9 \times 10^9$, $\Delta x=6500 \lambda$, and $h=4000$. }
  \label{fig5}
\end{figure}

Figure~\ref{fig5} presents the propagation of an ultra-intense LP wave along the magnetic field in the nonuniform magnetized plasma. At the beginning, when $b\lesssim a_0$, there is nearly no deformation and erosion since the interaction resembles the linear regime. At $t=3000\tau_0$, when the magnetic field drops to $b=30$ and the plasma density to $n_0=0.075n_\mathrm{c}$, the RCP component starts to erode from the pulse tail to its head. This counterintuitive erosion direction essentially results from the shift in the initial erosion point of the RCP component. As discussed above, the decreasing magnetic field will shift the initial erosion point from the tail to the head, while the decreasing density has a reverse effect. Consequently, the simulation here indicates that the decreasing magnetic field dominate on the shifting direction of the initial erosion point. After the pulse has propagated around $17.7\;\mathrm{km}$ at $t=15000\tau_0$, the RCP component is nearly fully depleted. The residual pulse exhibits an LCP mode with a CP degree of 93.7\%, and experiences only slight erosion in tenuous plasma. Therefore, the CP radio bursts generated by the MIAE effect should be able to travel out of the magnetosphere.

\section{Discussion and conclusion}
Most FRBs are LP modes. This can be attributed to the fact that the distributions of electrons and positrons are nearly symmetric in most regions of the magnetar magnetospheres.
CP bursts typically exhibit low or moderate CP degrees \cite{fengCircularPolarizationTwo2022,xuFastRadioBurst2022,choSpectropolarimetricAnalysisFRB2020,dayHighTimeResolution2020}. These partially CP bursts may result from the propagation in regions with a relatively high magnetic field or low plasma density [Fig.~\ref{fig4}]. Moreover, since a misalignment between the wave propagation and the magnetic field may also reduce the MIAE effect and CP degree, CP generation should be more efficient along the magnetic axis on the polar cap region, and the CP degree of FRBs may exhibit periodicity related to the magnetar’s spin.

Up to now, only four in the repeating FRBs---namely FRB 20121102A \cite{fengCircularPolarizationTwo2022}, FRB 20190520B \cite{fengCircularPolarizationTwo2022}, FRB 20201124A \cite{hilmarssonPolarizationPropertiesFRB2021,xuFastRadioBurst2022,kumarCircularlyPolarizedRadio2022,jiangNinetyPercentCircular2025}, and FRB 20220912A \cite{zhangFASTObservationsFRB2023,raviDeepSynopticArray2023,fengExtremelyActiveRepeating2024}---have shown CP modes. Fewer than $0.05\%$ of bursts from FRB 20121102A and FRB 20190520B exhibit CP modes \cite{fengCircularPolarizationTwo2022}, compared to over $16\%$ from FRB 20201124A and FRB 20220912A \cite{xuFastRadioBurst2022,zhangFASTObservationsFRB2023}. The lower CP proportions might be caused by higher magnetic fields in the sources, which suppress the MIAE effect. There are two supporting findings from the observations. First, for these repeating FRBs, the CP proportions and rotation measures ($\sim \int B_\mathrm{bg}(x)n(x) \mathrm{d}x$), which generally reflect the overall magnetic field strengths along the propagation paths, exhibit a negative correlation \cite{zhangFASTObservationsFRB2023}. Second, FRB 20121102A is hosted by a dwarf star-forming galaxy \cite{chatterjeeDirectLocalizationFast2017,tendulkarHostGalaxyRedshift2017,michilliExtremeMagnetoionicEnvironment2018}, and FRB 20190520B is associated with a persistent radio source \cite{niuRepeatingFastRadio2022}, suggesting they may originate from young magnetars owning stronger magnetic fields. In contrast, normal magnetars are more probable sources for FRB 20201124A and FRB 20220912A, given the significantly lower specific star formation rates in their host galaxies \cite{xuFastRadioBurst2022,raviDeepSynopticArray2023}.

To conclude, we mainly focus on the scenario that a nanosecond radio pulse propagates in a longitudinally magnetized electron-dominant plasma. When the pulse amplitude $a_0$ exceeds the background magnetic field $b$, we discover that the RCP and LCP components can undergo asymmetric erosion independently, leading to the generation of CP modes from an LP pulse. This asymmetry arises from the differing influence of the magnetic field on the nonlinearity of the wakefields excited by RCP and LCP pulses. We also demonstrate the CP generation of FRBs by applying this robust MIAE effect in magnetar magnetospheres. The final CP degree results from an entangled impact of various parameters, such as the pulse amplitude, the magnetic field strength, the plasma density, and the asymmetry between electrons and positrons. Future research will consider asymmetric pair plasmas with more realistic parameters in magnetar magnetospheres.

\begin{acknowledgments}
  We thank X.-L. Zhu for his help with EPOCH. This work was partially supported by the Strategic Priority Research Program of the Chinese Academy of Sciences (Grant No. XDA17040502). The PIC code EPOCH used here is funded by the UK EPSRC grants EP/G054950/1, EP/G056803/1, EP/G055165/1 and EP/ M022463/1.
\end{acknowledgments}

\bibliography{Refs}

\begin{thebibliography}{43}%
\makeatletter
\providecommand \@ifxundefined [1]{%
 \@ifx{#1\undefined}
}%
\providecommand \@ifnum [1]{%
 \ifnum #1\expandafter \@firstoftwo
 \else \expandafter \@secondoftwo
 \fi
}%
\providecommand \@ifx [1]{%
 \ifx #1\expandafter \@firstoftwo
 \else \expandafter \@secondoftwo
 \fi
}%
\providecommand \natexlab [1]{#1}%
\providecommand \enquote  [1]{``#1''}%
\providecommand \bibnamefont  [1]{#1}%
\providecommand \bibfnamefont [1]{#1}%
\providecommand \citenamefont [1]{#1}%
\providecommand \href@noop [0]{\@secondoftwo}%
\providecommand \href [0]{\begingroup \@sanitize@url \@href}%
\providecommand \@href[1]{\@@startlink{#1}\@@href}%
\providecommand \@@href[1]{\endgroup#1\@@endlink}%
\providecommand \@sanitize@url [0]{\catcode `\\12\catcode `\$12\catcode
  `\&12\catcode `\#12\catcode `\^12\catcode `\_12\catcode `\%12\relax}%
\providecommand \@@startlink[1]{}%
\providecommand \@@endlink[0]{}%
\providecommand \url  [0]{\begingroup\@sanitize@url \@url }%
\providecommand \@url [1]{\endgroup\@href {#1}{\urlprefix }}%
\providecommand \urlprefix  [0]{URL }%
\providecommand \Eprint [0]{\href }%
\providecommand \doibase [0]{https://doi.org/}%
\providecommand \selectlanguage [0]{\@gobble}%
\providecommand \bibinfo  [0]{\@secondoftwo}%
\providecommand \bibfield  [0]{\@secondoftwo}%
\providecommand \translation [1]{[#1]}%
\providecommand \BibitemOpen [0]{}%
\providecommand \bibitemStop [0]{}%
\providecommand \bibitemNoStop [0]{.\EOS\space}%
\providecommand \EOS [0]{\spacefactor3000\relax}%
\providecommand \BibitemShut  [1]{\csname bibitem#1\endcsname}%
\let\auto@bib@innerbib\@empty
\bibitem [{\citenamefont {Zhang}(2023)}]{zhangPhysicsFastRadio2023}%
  \BibitemOpen
  \bibfield  {author} {\bibinfo {author} {\bibfnamefont {B.}~\bibnamefont
  {Zhang}},\ }\href {https://doi.org/10.1103/RevModPhys.95.035005} {\bibfield
  {journal} {\bibinfo  {journal} {Rev. Mod. Phys.}\ }\textbf {\bibinfo {volume}
  {95}},\ \bibinfo {pages} {035005} (\bibinfo {year} {2023})}\BibitemShut
  {NoStop}%
\bibitem [{\citenamefont {Xiao}\ \emph {et~al.}(2021)\citenamefont {Xiao},
  \citenamefont {Wang},\ and\ \citenamefont {Dai}}]{xiaoPhysicsFastRadio2021}%
  \BibitemOpen
  \bibfield  {author} {\bibinfo {author} {\bibfnamefont {D.}~\bibnamefont
  {Xiao}}, \bibinfo {author} {\bibfnamefont {F.}~\bibnamefont {Wang}},\ and\
  \bibinfo {author} {\bibfnamefont {Z.}~\bibnamefont {Dai}},\ }\href
  {https://doi.org/10.1007/s11433-020-1661-7} {\bibfield  {journal} {\bibinfo
  {journal} {Sci. China Phys. Mech. Astron.}\ }\textbf {\bibinfo {volume}
  {64}},\ \bibinfo {pages} {249501} (\bibinfo {year} {2021})}\BibitemShut
  {NoStop}%
\bibitem [{\citenamefont {Andersen}\ \emph {et~al.}(2020)\citenamefont
  {Andersen} \emph {et~al.}}]{andersenBrightMilliseconddurationRadio2020}%
  \BibitemOpen
  \bibfield  {author} {\bibinfo {author} {\bibfnamefont {B.~C.}\ \bibnamefont
  {Andersen}} \emph {et~al.},\ }\href
  {https://doi.org/10.1038/s41586-020-2863-y} {\bibfield  {journal} {\bibinfo
  {journal} {Nature}\ }\textbf {\bibinfo {volume} {587}},\ \bibinfo {pages}
  {54} (\bibinfo {year} {2020})}\BibitemShut {NoStop}%
\bibitem [{\citenamefont {Kumar}\ \emph {et~al.}(2017)\citenamefont {Kumar},
  \citenamefont {Lu},\ and\ \citenamefont
  {Bhattacharya}}]{kumarFastRadioBurst2017}%
  \BibitemOpen
  \bibfield  {author} {\bibinfo {author} {\bibfnamefont {P.}~\bibnamefont
  {Kumar}}, \bibinfo {author} {\bibfnamefont {W.}~\bibnamefont {Lu}},\ and\
  \bibinfo {author} {\bibfnamefont {M.}~\bibnamefont {Bhattacharya}},\ }\href
  {https://doi.org/10.1093/mnras/stx665} {\bibfield  {journal} {\bibinfo
  {journal} {Mon. Not. Roy. Astron. Soc.}\ }\textbf {\bibinfo {volume} {468}},\
  \bibinfo {pages} {2726} (\bibinfo {year} {2017})}\BibitemShut {NoStop}%
\bibitem [{\citenamefont {Qu}\ and\ \citenamefont
  {Zhang}(2023)}]{quPolarizationFastRadio2023}%
  \BibitemOpen
  \bibfield  {author} {\bibinfo {author} {\bibfnamefont {Y.}~\bibnamefont
  {Qu}}\ and\ \bibinfo {author} {\bibfnamefont {B.}~\bibnamefont {Zhang}},\
  }\href {https://doi.org/10.1093/mnras/stad1072} {\bibfield  {journal}
  {\bibinfo  {journal} {Mon. Not. Roy. Astron. Soc.}\ }\textbf {\bibinfo
  {volume} {522}},\ \bibinfo {pages} {2448} (\bibinfo {year}
  {2023})}\BibitemShut {NoStop}%
\bibitem [{\citenamefont
  {Lyubarsky}(2014)}]{lyubarskyModelFastExtragalactic2014}%
  \BibitemOpen
  \bibfield  {author} {\bibinfo {author} {\bibfnamefont {Y.}~\bibnamefont
  {Lyubarsky}},\ }\href {https://doi.org/10.1093/mnrasl/slu046} {\bibfield
  {journal} {\bibinfo  {journal} {Mon. Not. Roy. Astron. Soc.: Letters}\
  }\textbf {\bibinfo {volume} {442}},\ \bibinfo {pages} {L9} (\bibinfo {year}
  {2014})}\BibitemShut {NoStop}%
\bibitem [{\citenamefont
  {Beloborodov}(2020)}]{beloborodovBlastWavesMagnetar2020}%
  \BibitemOpen
  \bibfield  {author} {\bibinfo {author} {\bibfnamefont {A.~M.}\ \bibnamefont
  {Beloborodov}},\ }\href {https://doi.org/10.3847/1538-4357/ab83eb} {\bibfield
   {journal} {\bibinfo  {journal} {Astrophys. J.}\ }\textbf {\bibinfo {volume}
  {896}},\ \bibinfo {pages} {142} (\bibinfo {year} {2020})}\BibitemShut
  {NoStop}%
\bibitem [{\citenamefont {Iwamoto}\ \emph {et~al.}(2024)\citenamefont
  {Iwamoto}, \citenamefont {Matsumoto}, \citenamefont {Amano}, \citenamefont
  {Matsukiyo},\ and\ \citenamefont
  {Hoshino}}]{iwamotoLinearlyPolarizedCoherent2024}%
  \BibitemOpen
  \bibfield  {author} {\bibinfo {author} {\bibfnamefont {M.}~\bibnamefont
  {Iwamoto}}, \bibinfo {author} {\bibfnamefont {Y.}~\bibnamefont {Matsumoto}},
  \bibinfo {author} {\bibfnamefont {T.}~\bibnamefont {Amano}}, \bibinfo
  {author} {\bibfnamefont {S.}~\bibnamefont {Matsukiyo}},\ and\ \bibinfo
  {author} {\bibfnamefont {M.}~\bibnamefont {Hoshino}},\ }\href
  {https://doi.org/10.1103/PhysRevLett.132.035201} {\bibfield  {journal}
  {\bibinfo  {journal} {Phys. Rev. Lett.}\ }\textbf {\bibinfo {volume} {132}},\
  \bibinfo {pages} {035201} (\bibinfo {year} {2024})}\BibitemShut {NoStop}%
\bibitem [{\citenamefont {Michilli}\ \emph {et~al.}(2018)\citenamefont
  {Michilli} \emph {et~al.}}]{michilliExtremeMagnetoionicEnvironment2018}%
  \BibitemOpen
  \bibfield  {author} {\bibinfo {author} {\bibfnamefont {D.}~\bibnamefont
  {Michilli}} \emph {et~al.},\ }\href {https://doi.org/10.1038/nature25149}
  {\bibfield  {journal} {\bibinfo  {journal} {Nature}\ }\textbf {\bibinfo
  {volume} {553}},\ \bibinfo {pages} {182} (\bibinfo {year}
  {2018})}\BibitemShut {NoStop}%
\bibitem [{\citenamefont {Cho}\ \emph {et~al.}(2020)\citenamefont {Cho} \emph
  {et~al.}}]{choSpectropolarimetricAnalysisFRB2020}%
  \BibitemOpen
  \bibfield  {author} {\bibinfo {author} {\bibfnamefont {H.}~\bibnamefont
  {Cho}} \emph {et~al.},\ }\href {https://doi.org/10.3847/2041-8213/ab7824}
  {\bibfield  {journal} {\bibinfo  {journal} {Astrophys. J. Lett.}\ }\textbf
  {\bibinfo {volume} {891}},\ \bibinfo {pages} {L38} (\bibinfo {year}
  {2020})}\BibitemShut {NoStop}%
\bibitem [{\citenamefont {Day}\ \emph {et~al.}(2020)\citenamefont {Day} \emph
  {et~al.}}]{dayHighTimeResolution2020}%
  \BibitemOpen
  \bibfield  {author} {\bibinfo {author} {\bibfnamefont {C.~K.}\ \bibnamefont
  {Day}} \emph {et~al.},\ }\href {https://doi.org/10.1093/mnras/staa2138}
  {\bibfield  {journal} {\bibinfo  {journal} {Mon. Not. Roy. Astron. Soc.}\
  }\textbf {\bibinfo {volume} {497}},\ \bibinfo {pages} {3335} (\bibinfo {year}
  {2020})}\BibitemShut {NoStop}%
\bibitem [{\citenamefont {Luo}\ \emph {et~al.}(2020)\citenamefont {Luo} \emph
  {et~al.}}]{luoDiversePolarizationAngle2020}%
  \BibitemOpen
  \bibfield  {author} {\bibinfo {author} {\bibfnamefont {R.}~\bibnamefont
  {Luo}} \emph {et~al.},\ }\href {https://doi.org/10.1038/s41586-020-2827-2}
  {\bibfield  {journal} {\bibinfo  {journal} {Nature}\ }\textbf {\bibinfo
  {volume} {586}},\ \bibinfo {pages} {693} (\bibinfo {year}
  {2020})}\BibitemShut {NoStop}%
\bibitem [{\citenamefont {Hilmarsson}\ \emph {et~al.}(2021)\citenamefont
  {Hilmarsson}, \citenamefont {Spitler}, \citenamefont {Main},\ and\
  \citenamefont {Li}}]{hilmarssonPolarizationPropertiesFRB2021}%
  \BibitemOpen
  \bibfield  {author} {\bibinfo {author} {\bibfnamefont {G.~H.}\ \bibnamefont
  {Hilmarsson}}, \bibinfo {author} {\bibfnamefont {L.~G.}\ \bibnamefont
  {Spitler}}, \bibinfo {author} {\bibfnamefont {R.~A.}\ \bibnamefont {Main}},\
  and\ \bibinfo {author} {\bibfnamefont {D.~Z.}\ \bibnamefont {Li}},\ }\href
  {https://doi.org/10.1093/mnras/stab2936} {\bibfield  {journal} {\bibinfo
  {journal} {Mon. Not. Roy. Astron. Soc.}\ }\textbf {\bibinfo {volume} {508}},\
  \bibinfo {pages} {5354} (\bibinfo {year} {2021})}\BibitemShut {NoStop}%
\bibitem [{\citenamefont {Kumar}\ \emph {et~al.}(2022)\citenamefont {Kumar},
  \citenamefont {Shannon}, \citenamefont {Lower}, \citenamefont {Bhandari},
  \citenamefont {Deller}, \citenamefont {Flynn},\ and\ \citenamefont
  {Keane}}]{kumarCircularlyPolarizedRadio2022}%
  \BibitemOpen
  \bibfield  {author} {\bibinfo {author} {\bibfnamefont {P.}~\bibnamefont
  {Kumar}}, \bibinfo {author} {\bibfnamefont {R.~M.}\ \bibnamefont {Shannon}},
  \bibinfo {author} {\bibfnamefont {M.~E.}\ \bibnamefont {Lower}}, \bibinfo
  {author} {\bibfnamefont {S.}~\bibnamefont {Bhandari}}, \bibinfo {author}
  {\bibfnamefont {A.~T.}\ \bibnamefont {Deller}}, \bibinfo {author}
  {\bibfnamefont {C.}~\bibnamefont {Flynn}},\ and\ \bibinfo {author}
  {\bibfnamefont {E.~F.}\ \bibnamefont {Keane}},\ }\href
  {https://doi.org/10.1093/mnras/stac683} {\bibfield  {journal} {\bibinfo
  {journal} {Mon. Not. Roy. Astron. Soc.}\ }\textbf {\bibinfo {volume} {512}},\
  \bibinfo {pages} {3400} (\bibinfo {year} {2022})}\BibitemShut {NoStop}%
\bibitem [{\citenamefont {Feng}\ \emph {et~al.}(2022)\citenamefont {Feng},
  \citenamefont {Zhang}, \citenamefont {Li}, \citenamefont {Yang},
  \citenamefont {Wang}, \citenamefont {Niu}, \citenamefont {Dai},\ and\
  \citenamefont {Yao}}]{fengCircularPolarizationTwo2022}%
  \BibitemOpen
  \bibfield  {author} {\bibinfo {author} {\bibfnamefont {Y.}~\bibnamefont
  {Feng}}, \bibinfo {author} {\bibfnamefont {Y.-K.}\ \bibnamefont {Zhang}},
  \bibinfo {author} {\bibfnamefont {D.}~\bibnamefont {Li}}, \bibinfo {author}
  {\bibfnamefont {Y.-P.}\ \bibnamefont {Yang}}, \bibinfo {author}
  {\bibfnamefont {P.}~\bibnamefont {Wang}}, \bibinfo {author} {\bibfnamefont
  {C.-H.}\ \bibnamefont {Niu}}, \bibinfo {author} {\bibfnamefont
  {S.}~\bibnamefont {Dai}},\ and\ \bibinfo {author} {\bibfnamefont {J.-M.}\
  \bibnamefont {Yao}},\ }\href {https://doi.org/10.1016/j.scib.2022.11.014}
  {\bibfield  {journal} {\bibinfo  {journal} {Sci. Bull.}\ }\textbf {\bibinfo
  {volume} {67}},\ \bibinfo {pages} {2398} (\bibinfo {year}
  {2022})}\BibitemShut {NoStop}%
\bibitem [{\citenamefont {Jiang}\ \emph {et~al.}(2025)\citenamefont {Jiang}
  \emph {et~al.}}]{jiangNinetyPercentCircular2025}%
  \BibitemOpen
  \bibfield  {author} {\bibinfo {author} {\bibfnamefont {J.}~\bibnamefont
  {Jiang}} \emph {et~al.},\ }\href {https://doi.org/10.1093/nsr/nwae293}
  {\bibfield  {journal} {\bibinfo  {journal} {Natl. Sci. Rev.}\ }\textbf
  {\bibinfo {volume} {12}},\ \bibinfo {pages} {nwae293} (\bibinfo {year}
  {2025})}\BibitemShut {NoStop}%
\bibitem [{\citenamefont {Feng}\ \emph {et~al.}(2024)\citenamefont {Feng} \emph
  {et~al.}}]{fengExtremelyActiveRepeating2024}%
  \BibitemOpen
  \bibfield  {author} {\bibinfo {author} {\bibfnamefont {Y.}~\bibnamefont
  {Feng}} \emph {et~al.},\ }\href {https://doi.org/10.3847/1538-4357/ad7a64}
  {\bibfield  {journal} {\bibinfo  {journal} {Astrophys. J.}\ }\textbf
  {\bibinfo {volume} {974}},\ \bibinfo {pages} {296} (\bibinfo {year}
  {2024})}\BibitemShut {NoStop}%
\bibitem [{\citenamefont {Lu}\ \emph {et~al.}(2020)\citenamefont {Lu},
  \citenamefont {Kumar},\ and\ \citenamefont
  {Zhang}}]{luUnifiedPictureGalactic2020}%
  \BibitemOpen
  \bibfield  {author} {\bibinfo {author} {\bibfnamefont {W.}~\bibnamefont
  {Lu}}, \bibinfo {author} {\bibfnamefont {P.}~\bibnamefont {Kumar}},\ and\
  \bibinfo {author} {\bibfnamefont {B.}~\bibnamefont {Zhang}},\ }\href
  {https://doi.org/10.1093/mnras/staa2450} {\bibfield  {journal} {\bibinfo
  {journal} {Mon. Not. Roy. Astron. Soc.}\ }\textbf {\bibinfo {volume} {498}},\
  \bibinfo {pages} {1397} (\bibinfo {year} {2020})}\BibitemShut {NoStop}%
\bibitem [{\citenamefont {Qu}\ \emph {et~al.}(2022)\citenamefont {Qu},
  \citenamefont {Kumar},\ and\ \citenamefont
  {Zhang}}]{quTransparencyFastRadio2022}%
  \BibitemOpen
  \bibfield  {author} {\bibinfo {author} {\bibfnamefont {Y.}~\bibnamefont
  {Qu}}, \bibinfo {author} {\bibfnamefont {P.}~\bibnamefont {Kumar}},\ and\
  \bibinfo {author} {\bibfnamefont {B.}~\bibnamefont {Zhang}},\ }\href
  {https://doi.org/10.1093/mnras/stac1910} {\bibfield  {journal} {\bibinfo
  {journal} {Mon. Not. Roy. Astron. Soc.}\ }\textbf {\bibinfo {volume} {515}},\
  \bibinfo {pages} {2020} (\bibinfo {year} {2022})}\BibitemShut {NoStop}%
\bibitem [{\citenamefont {Timokhin}\ and\ \citenamefont
  {Harding}(2015)}]{timokhinPOLARCAPCASCADE2015}%
  \BibitemOpen
  \bibfield  {author} {\bibinfo {author} {\bibfnamefont {A.~N.}\ \bibnamefont
  {Timokhin}}\ and\ \bibinfo {author} {\bibfnamefont {A.~K.}\ \bibnamefont
  {Harding}},\ }\href {https://doi.org/2017012522155900} {\bibfield  {journal}
  {\bibinfo  {journal} {Astrophys. J.}\ }\textbf {\bibinfo {volume} {810}},\
  \bibinfo {pages} {144} (\bibinfo {year} {2015})}\BibitemShut {NoStop}%
\bibitem [{\citenamefont {Philippov}\ \emph {et~al.}(2015)\citenamefont
  {Philippov}, \citenamefont {Cerutti}, \citenamefont {Tchekhovskoy},\ and\
  \citenamefont {Spitkovsky}}]{philippovINITIOPULSARMAGNETOSPHERE2015}%
  \BibitemOpen
  \bibfield  {author} {\bibinfo {author} {\bibfnamefont {A.~A.}\ \bibnamefont
  {Philippov}}, \bibinfo {author} {\bibfnamefont {B.}~\bibnamefont {Cerutti}},
  \bibinfo {author} {\bibfnamefont {A.}~\bibnamefont {Tchekhovskoy}},\ and\
  \bibinfo {author} {\bibfnamefont {A.}~\bibnamefont {Spitkovsky}},\ }\href
  {https://doi.org/10.1088/2041-8205/815/2/L19} {\bibfield  {journal} {\bibinfo
   {journal} {Astrophys. J. Lett.}\ }\textbf {\bibinfo {volume} {815}},\
  \bibinfo {pages} {L19} (\bibinfo {year} {2015})}\BibitemShut {NoStop}%
\bibitem [{\citenamefont {Wu}(2024)}]{wuMechanismCircularPolarization2024}%
  \BibitemOpen
  \bibfield  {author} {\bibinfo {author} {\bibfnamefont {H.-C.}\ \bibnamefont
  {Wu}},\ }\href {https://doi.org/10.3847/2041-8213/ad8154} {\bibfield
  {journal} {\bibinfo  {journal} {Astrophys. J. Lett.}\ }\textbf {\bibinfo
  {volume} {974}},\ \bibinfo {pages} {L21} (\bibinfo {year}
  {2024})}\BibitemShut {NoStop}%
\bibitem [{\citenamefont {Luo}\ and\ \citenamefont
  {Melrose}(2001)}]{luoCyclotronAbsorptionRadio2001}%
  \BibitemOpen
  \bibfield  {author} {\bibinfo {author} {\bibfnamefont {Q.}~\bibnamefont
  {Luo}}\ and\ \bibinfo {author} {\bibfnamefont {D.~B.}\ \bibnamefont
  {Melrose}},\ }\href {https://doi.org/10.1046/j.1365-8711.2001.04402.x}
  {\bibfield  {journal} {\bibinfo  {journal} {Mon. Not. Roy. Astron. Soc.}\
  }\textbf {\bibinfo {volume} {325}},\ \bibinfo {pages} {187} (\bibinfo {year}
  {2001})}\BibitemShut {NoStop}%
\bibitem [{\citenamefont {Wang}\ \emph {et~al.}(2010)\citenamefont {Wang},
  \citenamefont {Lai},\ and\ \citenamefont
  {Han}}]{wangPolarizationChangesPulsars2010}%
  \BibitemOpen
  \bibfield  {author} {\bibinfo {author} {\bibfnamefont {C.}~\bibnamefont
  {Wang}}, \bibinfo {author} {\bibfnamefont {D.}~\bibnamefont {Lai}},\ and\
  \bibinfo {author} {\bibfnamefont {J.}~\bibnamefont {Han}},\ }\href
  {https://doi.org/10.1111/j.1365-2966.2009.16074.x} {\bibfield  {journal}
  {\bibinfo  {journal} {Mon. Not. Roy. Astron. Soc.}\ }\textbf {\bibinfo
  {volume} {403}},\ \bibinfo {pages} {569} (\bibinfo {year}
  {2010})}\BibitemShut {NoStop}%
\bibitem [{\citenamefont {Dai}\ \emph {et~al.}(2021)\citenamefont {Dai} \emph
  {et~al.}}]{daiCircularPolarizationRepeating2021}%
  \BibitemOpen
  \bibfield  {author} {\bibinfo {author} {\bibfnamefont {S.}~\bibnamefont
  {Dai}} \emph {et~al.},\ }\href {https://doi.org/10.3847/1538-4357/ac193d}
  {\bibfield  {journal} {\bibinfo  {journal} {Astrophys. J.}\ }\textbf
  {\bibinfo {volume} {920}},\ \bibinfo {pages} {46} (\bibinfo {year}
  {2021})}\BibitemShut {NoStop}%
\bibitem [{\citenamefont {Goldston}\ and\ \citenamefont
  {Rutherford}(1995)}]{goldstonIntroductionPlasmaPhysics1995}%
  \BibitemOpen
  \bibfield  {author} {\bibinfo {author} {\bibfnamefont {R.~J.}\ \bibnamefont
  {Goldston}}\ and\ \bibinfo {author} {\bibfnamefont {P.~H.}\ \bibnamefont
  {Rutherford}},\ }\href@noop {} {\emph {\bibinfo {title} {Introduction to
  Plasma Physics}}}\ (\bibinfo  {publisher} {Institute of Physics Pub},\
  \bibinfo {address} {Bristol, UK ; Philadelphia},\ \bibinfo {year}
  {1995})\BibitemShut {NoStop}%
\bibitem [{\citenamefont {Bulanov}\ \emph {et~al.}(1992)\citenamefont
  {Bulanov}, \citenamefont {Inovenkov}, \citenamefont {Kirsanov}, \citenamefont
  {Naumova},\ and\ \citenamefont
  {Sakharov}}]{bulanovNonlinearDepletionUltrashort1992}%
  \BibitemOpen
  \bibfield  {author} {\bibinfo {author} {\bibfnamefont {S.~V.}\ \bibnamefont
  {Bulanov}}, \bibinfo {author} {\bibfnamefont {I.~N.}\ \bibnamefont
  {Inovenkov}}, \bibinfo {author} {\bibfnamefont {V.~I.}\ \bibnamefont
  {Kirsanov}}, \bibinfo {author} {\bibfnamefont {N.~M.}\ \bibnamefont
  {Naumova}},\ and\ \bibinfo {author} {\bibfnamefont {A.~S.}\ \bibnamefont
  {Sakharov}},\ }\href {https://doi.org/10.1063/1.860046} {\bibfield  {journal}
  {\bibinfo  {journal} {Phys. Fluids B: Plasma Physics}\ }\textbf {\bibinfo
  {volume} {4}},\ \bibinfo {pages} {1935} (\bibinfo {year} {1992})}\BibitemShut
  {NoStop}%
\bibitem [{\citenamefont {Decker}\ \emph {et~al.}(1996)\citenamefont {Decker},
  \citenamefont {Mori},\ and\ \citenamefont
  {Katsouleas}}]{deckerEvolutionUltraintenseShortpulse1996}%
  \BibitemOpen
  \bibfield  {author} {\bibinfo {author} {\bibfnamefont {C.~D.}\ \bibnamefont
  {Decker}}, \bibinfo {author} {\bibfnamefont {W.~B.}\ \bibnamefont {Mori}},\
  and\ \bibinfo {author} {\bibfnamefont {T.}~\bibnamefont {Katsouleas}},\
  }\bibfield  {journal} {\bibinfo  {journal} {Phys. Plasmas}\ }\textbf
  {\bibinfo {volume} {3}},\ \href {https://doi.org/10.1063/1.872001}
  {10.1063/1.872001} (\bibinfo {year} {1996})\BibitemShut {NoStop}%
\bibitem [{\citenamefont {Zhang}\ and\ \citenamefont
  {Wu}(2022)}]{zhangUpperFieldstrengthLimit2022}%
  \BibitemOpen
  \bibfield  {author} {\bibinfo {author} {\bibfnamefont {Y.}~\bibnamefont
  {Zhang}}\ and\ \bibinfo {author} {\bibfnamefont {H.-C.}\ \bibnamefont {Wu}},\
  }\href {https://doi.org/10.3847/1538-4357/ac5e2f} {\bibfield  {journal}
  {\bibinfo  {journal} {Astrophys. J.}\ }\textbf {\bibinfo {volume} {929}},\
  \bibinfo {pages} {164} (\bibinfo {year} {2022})}\BibitemShut {NoStop}%
\bibitem [{\citenamefont {Nimmo}\ \emph {et~al.}(2021)\citenamefont {Nimmo}
  \emph {et~al.}}]{nimmoHighlyPolarizedMicrostructure2021}%
  \BibitemOpen
  \bibfield  {author} {\bibinfo {author} {\bibfnamefont {K.}~\bibnamefont
  {Nimmo}} \emph {et~al.},\ }\href {https://doi.org/10.1038/s41550-021-01321-3}
  {\bibfield  {journal} {\bibinfo  {journal} {Nat. Astron.}\ }\textbf {\bibinfo
  {volume} {5}},\ \bibinfo {pages} {594} (\bibinfo {year} {2021})}\BibitemShut
  {NoStop}%
\bibitem [{\citenamefont {Nimmo}\ \emph {et~al.}(2022)\citenamefont {Nimmo}
  \emph {et~al.}}]{nimmoBurstTimescalesLuminosities2022}%
  \BibitemOpen
  \bibfield  {author} {\bibinfo {author} {\bibfnamefont {K.}~\bibnamefont
  {Nimmo}} \emph {et~al.},\ }\href {https://doi.org/10.1038/s41550-021-01569-9}
  {\bibfield  {journal} {\bibinfo  {journal} {Nat. Astron.}\ }\textbf {\bibinfo
  {volume} {6}},\ \bibinfo {pages} {393} (\bibinfo {year} {2022})}\BibitemShut
  {NoStop}%
\bibitem [{\citenamefont {Arber}\ \emph {et~al.}(2015)\citenamefont {Arber},
  \citenamefont {Bennett}, \citenamefont {Brady}, \citenamefont
  {{Lawrence-Douglas}}, \citenamefont {Ramsay}, \citenamefont {Sircombe},
  \citenamefont {Gillies}, \citenamefont {Evans}, \citenamefont {Schmitz},
  \citenamefont {Bell},\ and\ \citenamefont
  {Ridgers}}]{arberContemporaryParticleincellApproach2015a}%
  \BibitemOpen
  \bibfield  {author} {\bibinfo {author} {\bibfnamefont {T.~D.}\ \bibnamefont
  {Arber}}, \bibinfo {author} {\bibfnamefont {K.}~\bibnamefont {Bennett}},
  \bibinfo {author} {\bibfnamefont {C.~S.}\ \bibnamefont {Brady}}, \bibinfo
  {author} {\bibfnamefont {A.}~\bibnamefont {{Lawrence-Douglas}}}, \bibinfo
  {author} {\bibfnamefont {M.~G.}\ \bibnamefont {Ramsay}}, \bibinfo {author}
  {\bibfnamefont {N.~J.}\ \bibnamefont {Sircombe}}, \bibinfo {author}
  {\bibfnamefont {P.}~\bibnamefont {Gillies}}, \bibinfo {author} {\bibfnamefont
  {R.~G.}\ \bibnamefont {Evans}}, \bibinfo {author} {\bibfnamefont
  {H.}~\bibnamefont {Schmitz}}, \bibinfo {author} {\bibfnamefont {A.~R.}\
  \bibnamefont {Bell}},\ and\ \bibinfo {author} {\bibfnamefont {C.~P.}\
  \bibnamefont {Ridgers}},\ }\bibfield  {title} {\bibinfo {title} {Contemporary
  particle-in-cell approach to laser-plasma modelling},\ }\href
  {https://doi.org/10.1088/0741-3335/57/11/113001} {\bibfield  {journal}
  {\bibinfo  {journal} {Plasma Phys. Control. Fusion}\ }\textbf {\bibinfo
  {volume} {57}},\ \bibinfo {pages} {113001} (\bibinfo {year}
  {2015})}\BibitemShut {NoStop}%
\bibitem [{\citenamefont {Goldstein}(2011)}]{goldsteinPolarizedLight2011}%
  \BibitemOpen
  \bibfield  {author} {\bibinfo {author} {\bibfnamefont {D.~H.}\ \bibnamefont
  {Goldstein}},\ }\href@noop {} {\emph {\bibinfo {title} {Polarized
  {{Light}}}}},\ \bibinfo {edition} {third edition}\ ed.\ (\bibinfo
  {publisher} {CRC Press, Taylor \& Francis Group},\ \bibinfo {year}
  {2011})\BibitemShut {NoStop}%
\bibitem [{\citenamefont
  {Jackson}(1999)}]{jacksonClassicalElectrodynamics1999}%
  \BibitemOpen
  \bibfield  {author} {\bibinfo {author} {\bibfnamefont {J.~D.}\ \bibnamefont
  {Jackson}},\ }\href@noop {} {\emph {\bibinfo {title} {Classical
  {{Electrodynamics}}}}},\ \bibinfo {edition} {third edition}\ ed.\ (\bibinfo
  {publisher} {John Wiley \& Sons, INC.},\ \bibinfo {year} {1999})\BibitemShut
  {NoStop}%
\bibitem [{\citenamefont {Esarey}\ \emph {et~al.}(2009)\citenamefont {Esarey},
  \citenamefont {Schroeder},\ and\ \citenamefont
  {Leemans}}]{esareyPhysicsLaserdrivenPlasmabased2009}%
  \BibitemOpen
  \bibfield  {author} {\bibinfo {author} {\bibfnamefont {E.}~\bibnamefont
  {Esarey}}, \bibinfo {author} {\bibfnamefont {C.~B.}\ \bibnamefont
  {Schroeder}},\ and\ \bibinfo {author} {\bibfnamefont {W.~P.}\ \bibnamefont
  {Leemans}},\ }\href {https://doi.org/10.1103/RevModPhys.81.1229} {\bibfield
  {journal} {\bibinfo  {journal} {Rev. Mod. Phys.}\ }\textbf {\bibinfo {volume}
  {81}},\ \bibinfo {pages} {1229} (\bibinfo {year} {2009})}\BibitemShut
  {NoStop}%
\bibitem [{\citenamefont {Salamin}\ \emph {et~al.}(2000)\citenamefont
  {Salamin}, \citenamefont {Faisal},\ and\ \citenamefont
  {Keitel}}]{salaminExactAnalysisUltrahigh2000}%
  \BibitemOpen
  \bibfield  {author} {\bibinfo {author} {\bibfnamefont {Y.~I.}\ \bibnamefont
  {Salamin}}, \bibinfo {author} {\bibfnamefont {F.~H.~M.}\ \bibnamefont
  {Faisal}},\ and\ \bibinfo {author} {\bibfnamefont {C.~H.}\ \bibnamefont
  {Keitel}},\ }\href@noop {} {\bibfield  {journal} {\bibinfo  {journal} {Phys.
  Rev. A}\ } (\bibinfo {year} {2000})}\BibitemShut {NoStop}%
\bibitem [{\citenamefont {Lorimer}\ and\ \citenamefont
  {Kramer}(2005)}]{lorimerHandbookPulsarAstronomy2005}%
  \BibitemOpen
  \bibfield  {author} {\bibinfo {author} {\bibfnamefont {D.~R.}\ \bibnamefont
  {Lorimer}}\ and\ \bibinfo {author} {\bibfnamefont {M.}~\bibnamefont
  {Kramer}},\ }\href@noop {} {\emph {\bibinfo {title} {Handbook of {{Pulsar
  Astronomy}}}}}\ (\bibinfo  {publisher} {Cambridge University Press},\
  \bibinfo {year} {2005})\BibitemShut {NoStop}%
\bibitem [{\citenamefont {Xu}\ \emph {et~al.}(2022)\citenamefont {Xu} \emph
  {et~al.}}]{xuFastRadioBurst2022}%
  \BibitemOpen
  \bibfield  {author} {\bibinfo {author} {\bibfnamefont {H.}~\bibnamefont {Xu}}
  \emph {et~al.},\ }\href {https://doi.org/10.1038/s41586-022-05071-8}
  {\bibfield  {journal} {\bibinfo  {journal} {Nature}\ }\textbf {\bibinfo
  {volume} {609}},\ \bibinfo {pages} {685} (\bibinfo {year}
  {2022})}\BibitemShut {NoStop}%
\bibitem [{\citenamefont {Zhang}\ \emph {et~al.}(2023)\citenamefont {Zhang}
  \emph {et~al.}}]{zhangFASTObservationsFRB2023}%
  \BibitemOpen
  \bibfield  {author} {\bibinfo {author} {\bibfnamefont {Y.-K.}\ \bibnamefont
  {Zhang}} \emph {et~al.},\ }\href {https://doi.org/10.3847/1538-4357/aced0b}
  {\bibfield  {journal} {\bibinfo  {journal} {Astrophys. J.}\ }\textbf
  {\bibinfo {volume} {955}},\ \bibinfo {pages} {142} (\bibinfo {year}
  {2023})}\BibitemShut {NoStop}%
\bibitem [{\citenamefont {Ravi}\ \emph {et~al.}(2023)\citenamefont {Ravi} \emph
  {et~al.}}]{raviDeepSynopticArray2023}%
  \BibitemOpen
  \bibfield  {author} {\bibinfo {author} {\bibfnamefont {V.}~\bibnamefont
  {Ravi}} \emph {et~al.},\ }\href {https://doi.org/10.3847/2041-8213/acc4b6}
  {\bibfield  {journal} {\bibinfo  {journal} {Astrophys. J. Lett.}\ }\textbf
  {\bibinfo {volume} {949}},\ \bibinfo {pages} {L3} (\bibinfo {year}
  {2023})}\BibitemShut {NoStop}%
\bibitem [{\citenamefont {Chatterjee}\ \emph {et~al.}(2017)\citenamefont
  {Chatterjee} \emph {et~al.}}]{chatterjeeDirectLocalizationFast2017}%
  \BibitemOpen
  \bibfield  {author} {\bibinfo {author} {\bibfnamefont {S.}~\bibnamefont
  {Chatterjee}} \emph {et~al.},\ }\href {https://doi.org/10.1038/nature20797}
  {\bibfield  {journal} {\bibinfo  {journal} {Nature}\ }\textbf {\bibinfo
  {volume} {541}},\ \bibinfo {pages} {58} (\bibinfo {year} {2017})}\BibitemShut
  {NoStop}%
\bibitem [{\citenamefont {Tendulkar}\ \emph {et~al.}(2017)\citenamefont
  {Tendulkar} \emph {et~al.}}]{tendulkarHostGalaxyRedshift2017}%
  \BibitemOpen
  \bibfield  {author} {\bibinfo {author} {\bibfnamefont {S.~P.}\ \bibnamefont
  {Tendulkar}} \emph {et~al.},\ }\href
  {https://doi.org/10.3847/2041-8213/834/2/L7} {\bibfield  {journal} {\bibinfo
  {journal} {Astrophys. J. Lett.}\ }\textbf {\bibinfo {volume} {834}},\
  \bibinfo {pages} {L7} (\bibinfo {year} {2017})}\BibitemShut {NoStop}%
\bibitem [{\citenamefont {Niu}\ \emph {et~al.}(2022)\citenamefont {Niu} \emph
  {et~al.}}]{niuRepeatingFastRadio2022}%
  \BibitemOpen
  \bibfield  {author} {\bibinfo {author} {\bibfnamefont {C.-H.}\ \bibnamefont
  {Niu}} \emph {et~al.},\ }\href {https://doi.org/10.1038/s41586-022-04755-5}
  {\bibfield  {journal} {\bibinfo  {journal} {Nature}\ }\textbf {\bibinfo
  {volume} {606}},\ \bibinfo {pages} {873} (\bibinfo {year}
  {2022})}\BibitemShut {NoStop}%
\end{thebibliography}%

\end{document}